# Leveraging Urban Big Data for Informed Business Location Decisions: A Case Study of Starbucks in Tianhe District, Guangzhou City

Yan Xiang[1], Danni Chang[1], Xuan Feng[2]
[1]School of Design, Shanghai Jiao Tong University, Shanghai, China
[2]College of Engineering, Seoul National University, Seoul, South Korea
yanxiang@sjtu.edu.cn (Yan Xiang), dchang1@sjtu.edu.cn (Danni Chang)

*Abstract* - **With the development of the information age, cities provide a large amount of data that can be analyzed and utilized to facilitate the decision-making process. Urban big data and analytics are particularly valuable in the analysis of business location decisions, providing insight and supporting informed choices. By examining data relating to commercial locations, it becomes possible to analyze various spatial characteristics and derive the feasibility of different locations. This analytical approach contributes to effective decision-making and the formulation of robust location strategies. To illustrate this, the study focuses on Starbucks cafes in the Tianhe District of Guangzhou City, China. Utilizing data visualization maps, the spatial distribution characteristics and influencing factors of Starbucks locations are analyzed. By examining the geographical coordinates of Starbucks, main distribution characteristics are identified. Through this analysis, it explores the factors influencing the spatial layout of commercial store locations, using Starbucks as a case study. The findings offer valuable insights into the management of industrial layout and the location strategies of commercial businesses in urban environments, opening avenues for further research and development in this field.**

## I. INTRODUCTION

In the era of the information age, expanding companies in urban areas face the challenge of considering not only the economic and cultural disparities between cities but also the variations in business locations within cities [1]. And with the process of globalized consumption, multinational companies have accelerated their overseas expansion. However, there are few literature reports on the overseas location choice of multinational companies in the catering industry, represented by Starbucks [2]. While numerous studies have explored factors influencing commercial facility placement and Starbucks' sales strategies, insufficient attention has been given to its spatial distribution characteristics and influencing factors [3].

Such business and engineering management logic has been circulating in previous studies [4]. The commercial competitiveness and consumer dynamics of a region can be roughly measured based on the number of Starbucks coffee shops [5]. As the rate of store openings in China accelerates, Starbucks' city map will also become a more important reference for city strength. In previous hot discussions about Starbucks' spatial layout, Starbucks tends to concentrate on important transportation hubs and business districts [6]. In addition, urban centers, as composite centers providing political, economic and cultural functions, have influenced Starbucks' site selection to a certain extent. Meanwhile, as a consumer service company, Starbucks has weaker leasing capacity than restaurants and other businesses, and the cost of land price has an impact on its spatial layout [7].

Therefore, this paper aims to analyze commercial site selection data, taking Tianhe District in Guangzhou, China, as an example, and focuses on examining the spatial diffusion characteristics and influencing factors of Starbucks. This analysis endeavors to enhance decision-making processes and formulate effective site selection strategies, providing both theoretical and practical support for the management of commerce spatial layout and decision-making processes.

## II. METHODOLOGY

At the level of research methodology, we collect data by processing and visualizing them for different aspects of analysis, to derive the influence of different factors, as shown in Figure 1.

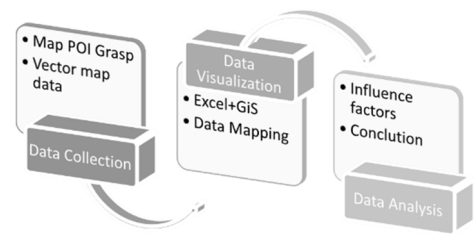

Fig. 1. Framework of research methodology.

*1) Data collection:* Accurate data collection is essential for conducting map visualization analysis. To ensure accuracy, we collected the precise geographical coordinates from latest months and real-time data of the distribution of Starbucks in Tianhe district. The use of a map POI coordinates grabber tool facilitates the collection of latitude and longitude data for the required elements, enabling accurate representation of their positions on the base map. This includes information on the surrounding land use, demographics, competitor locations, and infrastructure availability. Comprehensive dataset provides a more holistic understanding of the factors influencing business location decisions.

2) *Date visualization:* In terms of tools for data visualization, we use excel to visualize and present the data based on the type and kind of data we obtained. Excel provides a versatile platform that allows for effective representation and analysis of the data, the collected map coordinate data into the excel file can be processed quickly and obtain 3D numerical performance. At the same time, combined with GIS tools to achieve accurate visualization of location data. In addition to Excel, there are specialized data visualization software and GIS platforms that provide advanced mapping and spatial analysis capabilities. These tools enable the creation of interactive maps, heat maps, and 3D visualizations, allowing for a more comprehensive exploration of the relationships between different variables and their impact on business location decisions.

3) *Data analysis:* Taking Starbucks in Tianhe District as an example, we use big data visualization to analyze location. Due to the huge amount of data collected, we divide the data into three major categories: land price, human flow, and transportation:

*a.Land use &Land price:*

According to the nature of different land, select a number of representative target points in Tianhe District and calculate the average land price.

*b.Flow of people:*

Using Tencent's pedestrian traffic detection as a tool to record real-time traffic data. The tool collects video based on the embedded camera lens, then performs parallax calculation on the video images of the two cameras to form a 3D image of the person in the video, and then analyzes the shape and height of the human body, sets the statistics by region and direction to calculate the number of passengers.

*C. Transportation:*

The traffic information here is abstracted into public transportation coverage, including the radiation coverage density of bus stops and subway stations. The locations of relevant bus stops and subway stations are collected and analyzed for subsequent correlation analysis.

## III. RESULTS ANALYSIS

### A. Overall Spatial Distribution Analysis

1) *Spatial distribution status:*

According to the official data of Starbucks, there were 175 Starbucks coffee shops in Guangzhou, including 143 central urban areas (Liwan District, Yuexiu District, Haizhu District, Tianhe District, Baiyun District, Huangpu District), accounting for 81.7%, 26 in the suburbs (Panyu District, Huadu District), 6 in the remote suburbs (Nansha District, Zengcheng City, Conghua City), showing a rapid decline from the center.

Starbucks outlets exhibit a strong tendency to cluster in bustling urban hubs, such as city cores or central commercial zones. These locations commonly draw substantial pedestrian activity and provide a wide array of commercial and cultural amenities, as Figure 2 shows.

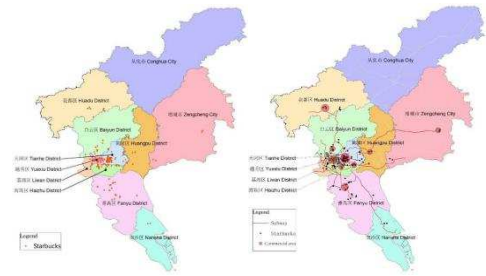

Fig. 2. The spatial distribution of Starbucks(left) and some major commercial areas(right) in Guangzhou.

Simultaneously, Starbucks strategically positions its establishments in close proximity to the prominent business hubs in Guangzhou, aligning with the well-developed public transportation network that facilitates the influx of a substantial consumer base. Tianhe District, situated in the heart of Guangzhou's downtown region, serves as a focal point for Starbucks' extensive presence within the city. The concentration of Starbucks stores predominantly centers around the core of Tianhe District, exhibiting a pronounced correlation with the patterns of public traffic flow (refer to Figure 3).

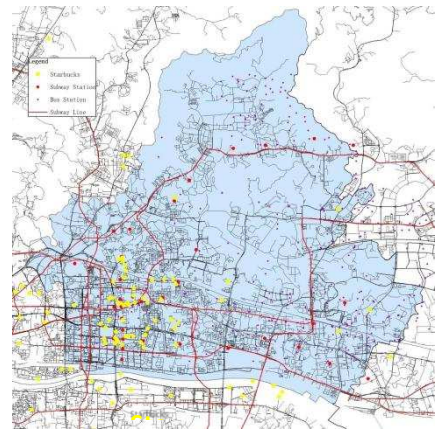

Fig. 3. The spatial distribution of Starbucks in Tianhe District of Guangzhou.

The distribution of Starbucks stores in Tianhe District reveals four primary typologies: street-facing storefronts on the first floor of large shopping malls, non-1st floor storefronts within large shopping malls, strip storefronts, and street-facing storefronts on the first floor of commercial office buildings, as well as non-1st floor street storefronts within commercial office buildings. Among these categories, street-facing storefronts on the first floor of large shopping malls account for 40% of the selected sites, occupying prime street-level spaces with high pedestrian footfall. Additionally, favorable locations include prominent business districts, strip commercial districts, and business office areas. It is noteworthy that only two Starbucks establishments are situated within the subterranean dining spaces of office buildings, as illustrated in TABLE I.

TABLE I
THE SPATIAL DISTRIBUTION STATUS OF STARBUCKS IN TIANHE DISTRICT

| Location Type | Location Diagram | Number (Proportion) |
| --- | --- | --- |
| Type-A: Street-facing storefront on 1st floor of large shopping malls | | 23(40.35%) |
| Type-B: Non-1F storefront in large shopping malls | | 10(17.54%) |
| Type-C: Strip storefront | | 13(22.81%) |
| Type-D: Street-facing storefront on 1st floor of commercial office buildings | | 9(15.79%) |
| Type-E: Non-first floor storefront in commercial office buildings | | 2(3.51%) |

*2) Spatial distribution characteristic:*

Based on a comprehensive analysis, the spatial distribution characteristics of Starbucks in the Tianhe District can be delineated by several key observations. Firstly, Starbucks stores are predominantly concentrated in the central area of Tianhe District, with a discernible decrease in density as one moves away from the center. Additionally, Starbucks exhibits a strong presence in prosperous regions, specifically within the business district and areas characterized by higher housing prices. Moreover, a significant correlation is observed between the Starbucks index and the density of bus and subway stations, indicating the influential role of public transportation availability in determining Starbucks locations. Lastly, Starbucks demonstrates a clear preference for densely populated areas, highlighting a strategic emphasis on regions with vibrant human activity and substantial foot traffic. By examining these spatial patterns and understanding the underlying factors, valuable insights can be gained for effective decision-making and the development of robust location strategies.

*B. Main influencing factors analysis*

*1) Land Use &Price:*

The distribution of commercial facilities and the development of economic activities are inherently tied to land availability. The level of land prices serves as a critical factor influencing the spatial layout of retail food and beverage establishments. Whether through land acquisition, leasing, or cooperative arrangements, the cost of land represents a significant portion of operating expenses for retail companies. Consequently, the spatial arrangement of retail enterprises demonstrates a certain sensitivity to land price levels. In this study, we quantitatively analyze the relationship between the spatial distribution of Starbucks stores and the benchmark land prices of commercial land in Guangzhou, as depicted in Figure 4.

Previous research has established that proximity to urban centers, well-developed transportation networks, and high pedestrian flows correspond to higher land price levels [8]. As one moves away from the city center, the land price level typically experiences a distance attenuation effect. The concentration of Starbucks stores near the central business district aligns with the areas characterized by higher commercial land price levels. As the distance from the business center increases, the level of commercial land prices decreases, resulting in a significant drop in the number of Starbucks stores. Hence, land price level stands as a pivotal factor influencing the spatial distribution of Starbucks stores. Additionally, the presence of high-end office buildings and upscale residential areas also plays a crucial role in site selection. These areas exhibit high levels of consumption and regional foot traffic, ensuring a stable customer base and purchasing power [9].

As a consumer-focused retail food company, Starbucks relies on the surrounding commercial atmosphere. Locations with vibrant commercial environments attract dense foot traffic, information flow, and logistics activities, exerting a positive influence on the development of retail enterprises. Areas such as shopping malls, high-end office buildings, and upscale residential areas that attract large crowds are given priority consideration during Starbucks' site selection process.

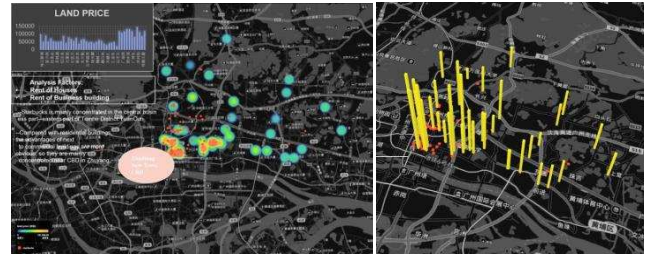

Fig. 4. The spatial distribution of starbucks with land price of Tianhe District of Guangzhou.

*2) Transportation:*

The accessibility of a city plays a crucial role in shaping the spatial arrangement of retail enterprises, serving as the vital connection between consumers and commercial facilities. Favorable traffic conditions not only save time and transportation costs for consumers but also enhance accessibility by concentrating a certain population density near retail establishments [10]. Thus, traffic conditions are a significant determinant influencing the selection of Starbucks store locations.

Our analysis revealed a significant correlation between the Starbucks Index and the density of bus and subway stations, confirming Starbucks' preference for areas with convenient transportation. The strategic selection of locations that capture high volumes of population movement along key routes is a critical factor influencing Starbucks' decision-making process, as depicted in Figure 5 and 6.

While the presence of Starbucks near bus stations is comparatively less pronounced than near subway stations. Bus stops are characterized by fast-paced pedestrian activity, whereas Starbucks establishments are designed for relaxation, leisure, and spending. Consequently, fewer Starbucks outlets are situated near bus stations.

The pronounced focus on public transportation hubs, particularly subway stations, underscores their pivotal role as primary gathering points for the city's population. Located along major passenger flow corridors, subway stations experience the highest influx of commuters per unit of time, thereby increasing the likelihood of attracting customers to Starbucks outlets. This factor contributes significantly to Starbucks' location decisions.

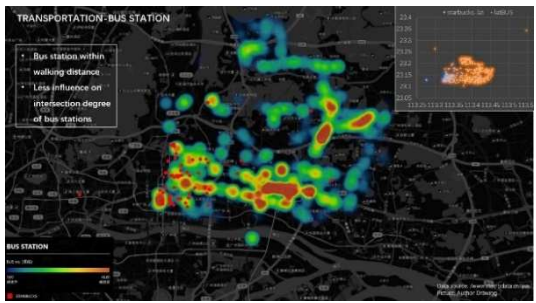

Fig. 5. The spatial distribution of Starbucks with bus station of Tianhe District of Guangzhou.

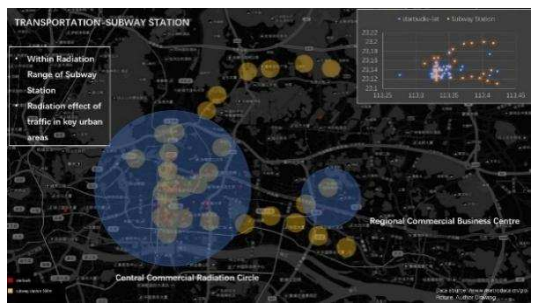

Fig. 6. The spatial distribution of Starbucks with subway station of Tianhe District of Guangzhou.

*3) Flow of Pedestrians:*

The relationship between Starbucks' commercial location and pedestrian flow is a crucial aspect in understanding the impact of foot traffic on Starbucks' site selection. Areas with dense pedestrian flow offer greater visibility and exposure to potential customers, increasing the likelihood of attracting walk-in customers and generating higher sales [11].

By analyzing the correlation between Starbucks' commercial locations and pedestrian flow, it becomes possible to identify patterns and preferences in site selection. This analysis involves considering factors such as busy shopping districts, transportation hubs, office complexes, universities, and other areas where people congregate and traverse frequently.

Based on our real-time data collection and visual analysis, it is evident that Starbucks exhibits a preference for locations characterized by bustling shopping malls, dense office complexes, and upscale residential areas. These areas are known for their popularity, stability, significant foot traffic, and prolonged customer visits. The flow of people plays a critical role in business location decisions, and Starbucks strategically chooses locations with high pedestrian traffic, as illustrated in Figure 7.

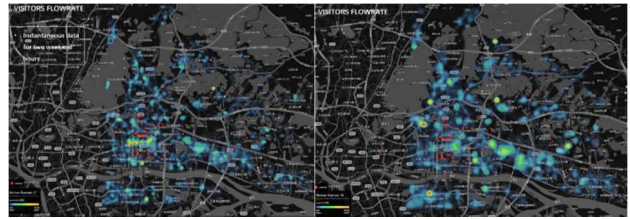

Fig. 7. The spatial distribution of Starbucks with pedestrians's flowrate in Tianhe District of Guangzhou

## IV. DISCUSSION

Through the above analysis, we found that there are several factors that influence the choice of location, and concluded different spatial distribution characteristics, which can be discussed specifically from the following:

The spatial distribution characteristics of Starbucks refer to the patterns and arrangement of Starbucks locations in a given geographical area. These characteristics provide insights into how Starbucks chooses its store locations and the factors that influence their distribution. While the specific spatial distribution characteristics may vary depending on the region or city under study, our study yielded some generalized findings.

*Concentration in urban centers:* Starbucks stores tend to be highly concentrated in urban centers, such as city downtowns or central business districts. These areas typically attract a large volume of foot traffic and offer a diverse range of commercial and cultural activities.

*Proximity to transportation hubs:* Starbucks often selects locations near transportation hubs, such as train stations, bus terminals, or major intersections. These areas serve as convenient meeting points and attract commuters, travelers, and individuals passing through the vicinity. The concentration of hot spots near Guangzhou East Railway Station stands out as the primary area. Serving as a crucial transportation hub, Guangzhou East Railway Station holds significant importance as a passenger distribution center in Guangzhou. Additionally, it serves as an interchange station for Guangzhou Metro Lines 1 and 3, while also being a major gathering point for Guangzhou-Shenzhen Hexie-Train and Metro Airport Express. Consequently, its well-developed transportation infrastructure and high population density establish it as a prominent location for Starbucks' spatial distribution.

*Presence in high-traffic areas:* Starbucks stores are commonly found in high-traffic areas with a significant number of pedestrians, including busy shopping streets, commercial complexes, and popular tourist destinations. These locations provide increased visibility and accessibility to a larger customer base.

*Clustered in commercial and retail zones:* Starbucks tends to cluster its stores within commercial and retail zones, where there is a concentration of other businesses, shops, and entertainment venues. This clustering strategy allows Starbucks to benefit from the existing customer flow and capitalize on the synergy created by neighboring establishments. The primary hotspots for Starbucks space distribution in Guangzhou are concentrated in two main areas: Tianhe City business district and Pearl River New City business district. These locations, characterized by vibrant commercial atmospheres and high footfall, have emerged as significant hotspots for Starbucks' spatial presence. Hotspot analysis reveals that Starbucks stores are prominently located in the city's commercial districts, areas with heavy traffic, and gathering spots for high-end consumer groups.

Through analysis, it can be seen that the city's urban centers, transportation hubs, high-traffic areas, and commercial and retail zones of a city are the main factors affecting the location of Starbucks. It is important to acknowledge that these spatial distribution characteristics may vary depending on the specific context and market conditions. Nevertheless, this paper's examination of the spatial distribution of Starbucks in a particular area offers valuable insights into their site selection strategy and the factors that influence their location decisions.

## V. CONCLUSION

This study focuses on analyzing the affect factors of spatial distribution of Starbucks stores in Tianhe District, Guangzhou City, China, as the research subjects. Spatial statistics are employed to investigate the spatial distribution characteristics of Starbucks and the factors influencing location selection.

Our analysis of Starbucks in Guangzhou reveals a significant spatial pattern. Starbucks stores are distributed in a compact band with a hierarchical structure, showing reduced density from the city center outward. Key factors driving these location choices include proximity to transportation networks and prominent business districts, ensuring accessibility for residents and office-goers. High-traffic areas are favored, aligning with Starbucks' appeal as a social coffee destination. Additionally, cost considerations, as seen in land prices, play a role in selecting optimal locations. Starbucks, like any business, will carefully weigh the costs associated with leasing or purchasing property against the potential revenue generated by each location.

Overall, this study not only offers valuable insights into the spatial dynamics of Starbucks in Guangzhou but also provides a framework for understanding how multinational companies strategically position themselves in urban environments to meet the evolving demands of their customers [12]. These insights can be instrumental in guiding future business expansion and urban planning efforts in similar contexts. Furthermore, engaging in additional theoretical discourse and practical validations will further bolster the continual advancement of this research area.